\documentclass[aps,twocolumn,showpacs]{revtex4}
\usepackage{graphicx}%
\usepackage{dcolumn}
\usepackage{amsmath}
\def\MSA{$\mu$-\textsc{squid}~}
\def\MSAs{$\mu$-\textsc{squids}~}

\def\MSB{$\mu$-\textsc{squid}}

\begin{document}
%\draft

\title{Scanning \MSA Force Microscope}
\author{C. Veauvy}
\affiliation{ CRTBT-CNRS, associ\'{e} \`{a}
l'Universit\'{e} Joseph Fourier \\ 25 avenue des Martyrs,  BP 166 X, 38042
Grenoble, France}
\author{D. Mailly}
\affiliation{LPN-CNRS \\ 196 H.
Ravera, 92220 Bagneux, France}
\author{K. Hasselbach}
\affiliation{
CRTBT-CNRS, associ\'{e} \`{a} l'Universit\'{e} Joseph Fourier \\ 25
avenue des Martyrs, BP 166 X, 38042 Grenoble, France}
\date{\today}

\begin{abstract}
 A novel scanning probe technique is presented: Scanning \MSA Force
 microscopy (SSFM). The instrument features independent topographic and
 magnetic imaging.  The SSFM operates in a dilution refrigerator in
 cryogenic vacuum.  Sample and probe can be cooled to 0.45 K. The probe
 consists of a \MSA placed at the edge of a silicon chip attached to a
 quartz tuning fork.  A topographic vertical resolution of 0.02 $\mu m$
 is demonstrated and magnetic flux as weak as $10^{-3} \Phi_{0}$ is
 resolved with a 1 $\mu m$ diameter \MSA loop.
\end{abstract}

\maketitle
%07.79.-v|Scanning probe microscopes and components
%near-field scanning, 07.79.F
%Piezoelectric devices, 85.50
%SQUID devices, 85.25.D 
%74.25.Ha|Magnetic properties of superconductors

%07.79.-v|Scanning probe microscopes and components
%near-field scanning, 07.79.F
%Piezoelectric devices, 85.50
%SQUID devices, 85.25.D 
%74.25.Ha|Magnetic properties of superconductors

\section{INTRODUCTION}

 A recent review \cite{Bending_1} on local magnetic
probes describes different techniques used for room temperature as
well as for low temperature magnetic imaging.  Several important
points have to be considered conceiving a scanning magnetic probe
microscope: Magnetic sensitivity and spatial resolution are determined
by the probe, the sample and the distance between them. It has been
observed \cite{Bending_1} that the sensitivity to the magnetic field, in
units of tesla, increases with probe size, but at the same time spatial
resolution diminishes.  The distance between sample and probe must be as
small as possible in order to improve the magnetic coupling between
them.  Therefore the control of this distance  becomes an important aspect for a
scanning magnetic probe microscope (SMPM). The choice of probe and of
distance control depends on the applications of the microscope.  The
main application of our microscope is the observation of vortices in
superconductors at low temperature.

In a superconductor each vortex
carries a quantized flux of $\Phi_{0}$= h/2 e = $2\cdot 10^{-15}$ Wb. The
magnetic size of a vortex is given by the penetration length,
$\lambda$, setting the length scale over which the screening currents
circulate around the normal vortex core. The typical penetration length is
of the order of 100 nm at low temperature, it increases with temperature,
to finally diverge at $T_{c}$.  The spatial resolution of a
SMPM imaging vortices can hardly be better than $\lambda$.
The probe should be of size $\lambda$ or smaller
and the scan height less than $\lambda$ in order to avoid convolution effects.
The distance between probe and sample must be controlled using a
technique compatible with the cryogenic environment, heating neither
sample nor tip.

 MFM microscopes have been built for imaging vortices in superconductors
 at temperatures as low as 4.2 K using either optical \cite{Hug} or
 piezoresistive \cite{Haesendonk} detection schemes. Scanning Hall
 probe microscopy \cite{Bending} images vortices at temperatures above
 5 K with a STM based technique to control the approach.  Very
 successful SQUID imaging has been done using mechanical translation
 without a separate approach control \cite {John}  -
 \cite{vanHarlingen}.  A scanning SQUID susceptometer \cite{Moler} has
 also been presented, using a piezoelectric scanner and a capacitive
 technique to control the approach.

 %The spatial resolution is
% limited by the size of the SQUID's pick-up loop 8 x 8 $\mu m^{2}$ and the
% distance between tip and sample.
 
%(John Wellstood Clark japonais van Harlingen). 
%Other groups combined scanning probe microscopy and magnetic imaging.
%For temepratures below 4.2 K only 
%very few magnetic imaging techniques exist.
%For applications at low temperature care must be taken to limit the 
%heating due to the feed-back. 
%MFM imaging at low temperatures uses either an optical readout 
%system (4.2 K) \cite{Hug}, a piezoresistive cantilever \cite{Haesendonk}, or 
%the tunneling current \cite{Bending}.
In  several near field microscopes
\cite{Schweinbock}-\cite{Grober} the piezoelectric effect has been used
recently to measure the minute forces between tip and sample.  The tip
is attached to the piezoelectric element, and the assembly is excited
at its mechanical resonance.  Upon approach, due to the surface
forces, the resonance frequency increases and the amplitude of
oscillation decreases. Shear forces are sensed when the tip oscillates at
constant height parallel to the sample's surface, and contact forces
appear when the tip oscillates perpendicular to the surface.
% A room temperature scanning Hall probe
%microscope \cite{Schweinbock} with shear force distance control has
%been reported.
Near field scanning optical microscopes have been built \cite{Karrei},
taking advantage of the high quality factor of piezoelectric quartz
crystals.  An optical fiber is attached to one prong of a quartz
tuning fork and the shear forces are measured.  Even a small STM tip
\cite{Giessibl}, \cite{Rychen_1999} can be attached to the tuning fork
in order to combine AFM and STM techniques in a contact force
geometry. It has been shown that this detection scheme can be used at
temperatures as low as 2.2 K \cite{Rychen_2000} and in high magnetic
fields.

We built a scanning \MSA force microscope (\MSA-FM) operating inside a
dilution refrigerator.  A quartz tuning fork senses the contact forces
between the tip of the \MSA chip and the sample's surface.  The
instrument has high magnetic spatial resolution, works at temperatures
as low as 0.45 K and is capable of acquiring simultaneously
topographic and magnetic images.  This microscope is to our knowledge
the first magnetic probe microscope working at such low temperatures.
% The optical fiber is attached to
%a tuning fork, the change of amplitude or shift of resonnance frequency
%upon approach are used to keep the distance between sample and tip
%constant, resulting in spatially resolved images

\section{MICROSCOPE SETUP}
\begin{figure} %[h!tb] %h=here,!=au plus prs,t=top,b=bottom
%\begin{figure}[c]
%	\includegraphics[width=8cm]{DESSIN_FM.eps}
    \includegraphics*[width=8cm]{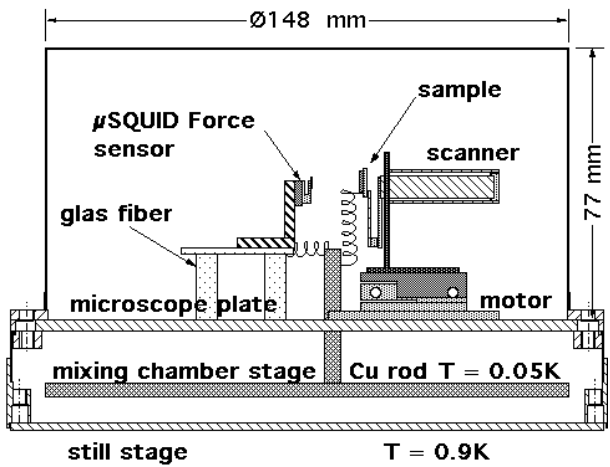}
	\caption{Schematic drawing of the Scanning \MSA Force Microscope inside the
	reversed dilution refrigerator.  The microscope plate is mounted on
	the screens of the still stage, sample and \MSA Force sensor are
	thermalized at the mixing chamber though silver foil and a copper.}
	\label{fig_setup}
\end{figure}

%moteur lineaire
%mesure du dŽplacement
%efficacitŽ chaud froid.
%Montage dans un cryostat renversŽ
%thermalisation sur B1
%thermalisation separŽ du SQUID
%dessin
The reversed dilution refrigerator \cite{Air_Liquide} is inside a cylindrical vacuum
chamber of 400 mm height and 200 mm diameter.  The refrigerator's base
temperature is 25 mK with 30 $\mu$W of cooling power at 100 mK. The
refrigerator is anchored to a concrete slab mounted on air suspension and a
helium dewar is attached underneath, feeding the 4.2 K pot of the
refrigerator by a continuous flow of liquid He.  Still and mixing
chamber are above the 4.2 K stage. 
The magnetic field is applied by external Helmholtz coils. The SSFM is
mounted on a copper plate fixed to a screen of the dilution
refrigerator. On the plate, Fig.  \ref{fig_setup}, is a linear
piezoelectric motor carrying a large range scanner \cite{Field} with
the sample.  In front of the scanned sample is the immobile \MSA Force
sensor.  The microscope plate, Fig.  \ref{fig_setup}, is either
flanged to copper screens of the 4.2 K stage or - of the still stage,
stages with sufficient cooling power to absorb the energy dissipated
during the coarse approach.  The sensor is mounted on three glass
fiber tubes, isolating it thermally from the plate.  The sample is
weakly coupled to the main body of the microscope as the ceramic
pieces of the scanner have poor thermal conduction.  Silver foil
thermal links connected to the mixing chamber through a copper rod,
allow us to cool sample and \MSA. Sample and \MSA reach a base
temperature of 0.45 K while the still is at 0.9 K.

The linear motor is
made of titanium and driven by four piezoelectric tubes. The piezoelectric
tubes are controlled by an arbitrary wave form generator whose output
is amplified by (APEX PA15) amplifiers. The voltage amplitudes can be
as high as 200 Volt. We incorporated a capacitive displacement sensor
between the base and the mobile part of the motor, allowing us to
follow in situ the motor's displacement.  The sensor is based on the principle
of a three terminal capacitive bridge used for dilatometers
\cite{Tremolet} or displacement sensors \cite {Kolb}.
%Our sensor consists of two plates, each carries photo lithographically
%defined interdigitated electrodes of 1.3 mm period. The two
%electrodes of the base are supplied by 180 degree out of phase
%AC-Voltage of 10 V amplitude. A voltage is induced in the upper
%electrodes and is read by a lockin-amplifier. The bridge is nulled
%when the upper electrode is above (0.1 mm gap) and in the middle of
%two adjacent base electrodes. 
The sensor has a resolution of the order of 10 nm.

\section{PROBE}

%on commence avec le SQUID
% taille etc\ldots..
%The purpose of the microscope is to observe vortex lattices in bulk 
%superconductors and artificial wire networks at low temperatures. 
%

In order to obtain high spatial magnetic resolution with the microscope we
chose as detector a 1 $\mu m$ diameter DC-SQUID, a \MSB. 
The Josephson Junctions of these \MSAs are composed of
Dayem bridges of 300 nm length, 20 nm width and 30 nm thickness. The
width of a \MSA branch is 200 nm.
Advantages of these \MSAs are numerous: they are easily manufactured
(only one single level of
e-beam lithography is necessary) and the flux pinning is weak as the
line width is very narrow. A magnetic flux change is detected by the 
SQUID through a change in critical current: the \MSA is biased with a 
DC-current ramp  which is reset to zero when the
critical current is reached and the \MSA becomes superconducting
again  \cite{Physica_C}, \cite{Chapelier}. The duration of the ramp is measured, it is proportional to
the critical current of the \MSB. The ramp is repeated every 1.6 ms.
The \MSAs made of Nb are
used for "high" temperature or high field applications but for high resolution
measurements Al \MSAs are preferable \cite{Physica_C}.  The
magnetization reversal of a 25 nm diameter ferromagnetic cobalt
particle \cite{Wolfgang} has been studied using a \MSB. The particle
was placed on the $\mu$-bridge.  This is the place for optimal flux
linkage between a very small flux source and the \MSB. Elsewhere most
of the particle's flux closes \cite{Physica_C} upon itself before it
intersects the \MSB.

\begin{figure}%[h!tb] %h=here,!=au plus prs,t=top,b=bottom

\includegraphics*[width=4cm]{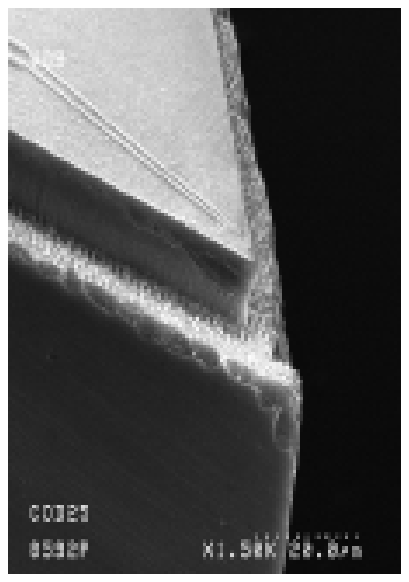}
\includegraphics*[width=4cm]{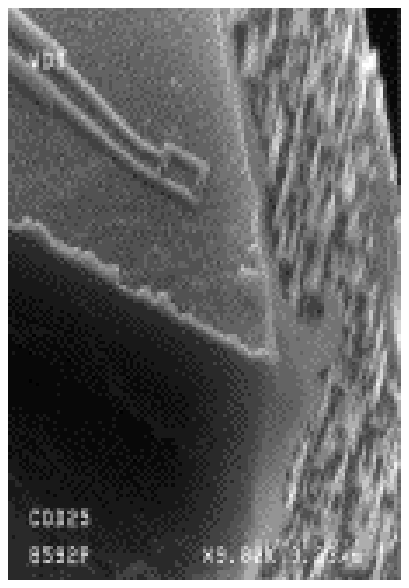}

\caption{SEM pictures of the \MSA sensor.  left:
Overview, the bulk Si is cut with a dicing machine at an angle of 25
degrees.  Current leads and the small \MSA loop are visible on the
mesa.  right: The \MSA sensor of 1 $\mu $m diameter is closely aligned
to the etched mesa, the $\mu$-bridge junctions can be distinguished. 
}
	\label{pointe}
\end{figure}

This example has illustrated the importance of proximity between sample 
and probe. For microfabricated probes the
objective 
of high spatial resolution can only be achieved if they are placed on 
the very edge of their support and if an appropriate distance control 
is used. First we describe our probe fabrication and then we present 
the distance control employed.  50 probes are fabricated on a 2 inches
diameter Si wafer.  Each \MSA chip is precision cut \cite{Gambarini}
with a dicing machine.  A precision of 5 $\mu m$ is obtained, limited
only by chipping of the Si.  The typical distances between tip and
\MSA is between 5 and 15 $\mu m$.  The distance between \MSA and
sample can be reduced further if the point of contact is closer to the
\MSB. This can be obtained by dry etching the Si substrate: The \MSA
is covered by an insulator ($SiC$ or $Si{_{3}}N{_{4}}$) and then a
PMMA mask is fabricated with the JEOL 5DIIU electron-beam writer
leaving open only the \MSB. A thick Al film is deposited and the
resist mask is then lifted, leaving an Al mask on the \MSB. The
following RIE etch \cite{Ayela} removes vertically 5 to 10 $\mu m$ of
the insulating layer and the unprotected Si.  The Al mask is removed
by wet etching.  The \MSA is situated at the edge of a mesa, Fig. 
\ref{pointe}.  The etched apex close to the \MSA serves now as tip for
distance control and topographic imaging.  The distance between \MSA
and tip is reduced to 2 or 3 $\mu $m.  For a 3 $\mu m$ distance
between tip and \MSA and an approach angle of 5 degrees the distance
between \MSA and sample is 0.26 $\mu $m.

 \begin{figure}%[h!tb]
 \includegraphics*[width=3.5cm]{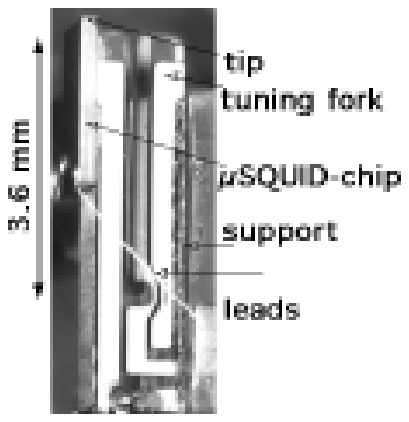}
\includegraphics*[width=5.5cm]{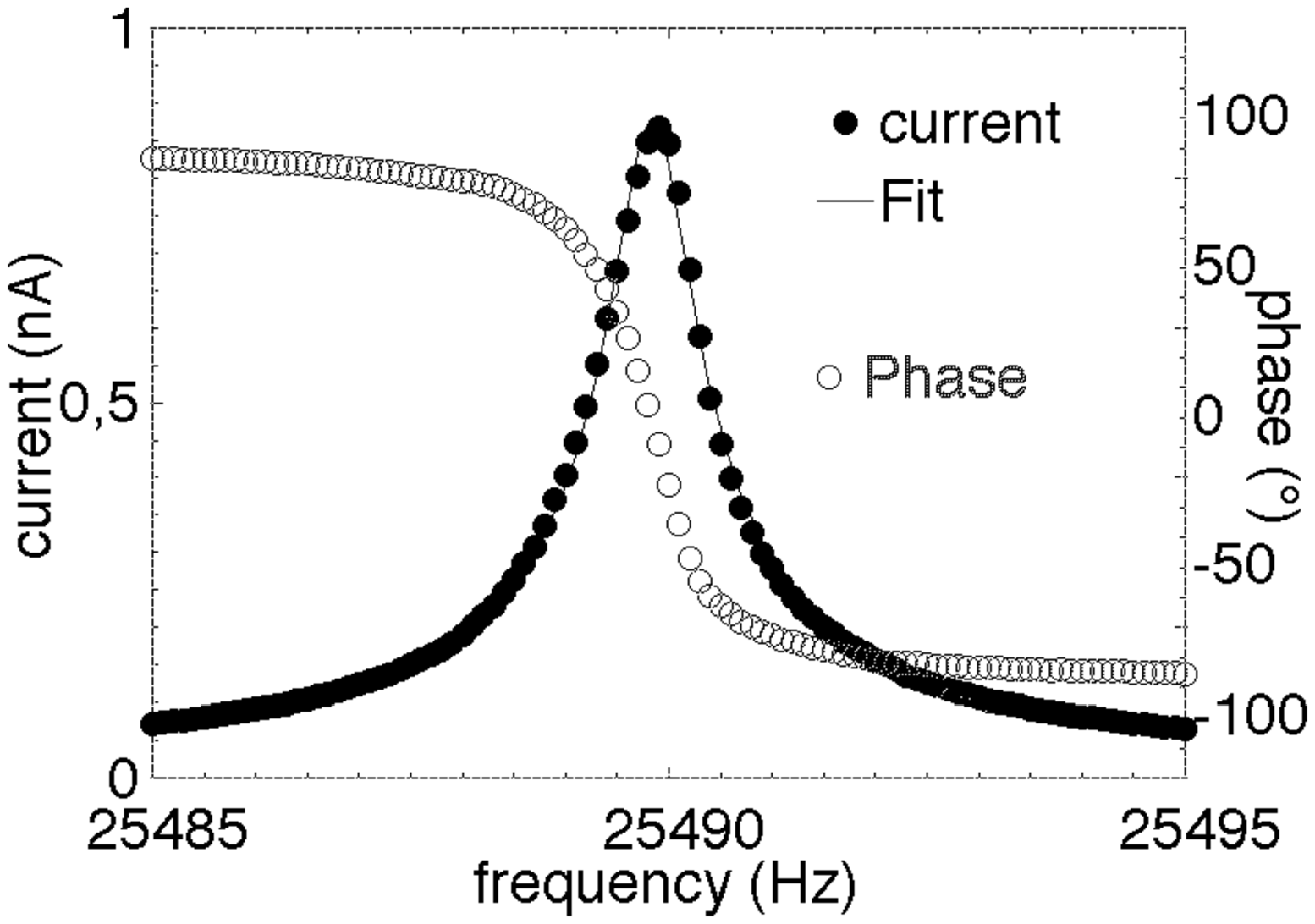}
\caption{left: the tuning fork carrying the \MSB-chip.  One side of the
tuning fork is attached to a fiberglass support.  The sensor is
oriented nearly parallel to the sample's surface.  right: Phase and amplitude
measured at 0.45 K of a tuning fork carrying a \MSB-chip, excited
mechanically at resonance.  The quality factor of the resonance is
17700.  The line corresponds to the fit, equation~$\ref{one}$.}
 \label{diapason}
\end{figure}

The proximity between \MSA and sample needs a good control of the
 tip-sample distance.  Three choices are possible: either tip and
 sample are continuously in contact or a distance sensor is incorporated
 and a feedback loop is used, or scanning takes place at constant
 height above the sample's surface.  We have chosen to control distance
 with a detector sensitive to surface forces.  The force sensor should
 have power consumption in the nW range in order to be compatible with
 low temperature applications and to be easily adapted to a dilution
 refrigerator.  We use a tuning fork distance control for the scanning
 \MSA Force microscope allowing us to separate the \MSA sensor
 development and the distance control.  The tuning fork is a 6 mm long,
 0.35 mm wide quartz watch crystal.  Each tine is 3.7 mm long and 0.63
 mm high.  It is designed for a resonance frequency of 32765 Hz at room
 temperature.  Scanning probe schemes \cite {Karrei}-\cite{Grober} were
 developed using tuning forks for near field scanning optical
 microscopy or AFM.

 %h=here,!=au plus prs,t=top,b=bottom
%\begin{figure}[c]
%diapason_SQUID_72_dpi_2.eps

In our case the \MSB-chip of 2.5 mm length weighs about 30 ${\%}$ of
the mass of one prong and electrical connections to the \MSA have to be
made. We choose to epoxy one prong to the support while the other
prong carries the Si chip on which the \MSA is fabricated. The chip
is glued on the prong and covers about two third of the free length of the
tine, Fig.  \ref{diapason} left.  Two AlSi wires of 33 $\mu
$m diameter are wedge-bonded onto the chip to connect the \MSA
electrically.  The tuning fork is mechanically excited with a
thickness mode piezoelectric crystal nearby and the generated current
is amplified at room temperature with a preamplifier of gain $35 \cdot
10^{-9}$ A/V.

The resonance frequency of the tuning fork is shifted from 32.7 kHz to
about 25 kHz by the extra mass of the Si chip.  Knowing the Young's
modulus of quartz, and using the expression of the spring constant of
a beam of given dimensions, a spring constant of the order of 36000
N/m is obtained for the bare tine.  The frequency variation of the
electrical current, $I(f')$, through the tuning fork can be fitted by
the Lorentz form of a serial (RLC) resonator as expressed in equation
\ref{one}.  The quality factor of the resonator is defined as $Q=\frac
{f} {\Delta f}$, with $\Delta f$ the width of the resonance at half
height.  The Q-factor for our resonators is of the order of 100 to 200
at room temperature in air and 10000 to 20000 at 0.45 K.
%h_{z}(\mathbf{r},z)= \frac{\Phi_{0}}{(2 \pi \lambda_{ab})^{2}
%}$$I{f}= \frac{f}{$\sqrt{f^{2}+3 Q^{2} (1-f^{2})^{2} } } $$$$I(f)=
%\frac{f}{1} $$

\begin{equation}
I(f{^\prime})= \frac{I_{0}f{^\prime}} {\sqrt{f{^\prime}^{2} +3 Q^{2}
(1-f{^\prime}^{2})^{2} }} + I_{base}
\label{one}
\end{equation}

%\sqrt{ f'^{2} +3 Q^{2} (1-f'^{2})^{2} } 

where, $f{^\prime}=\frac{f}{f_{res}}$ is the reduced frequency and $I_{0}$ is
the current amplitude at resonance.  The quality of the adjustment
indicates that the mechanical properties of our probe can be described
by this simple model despite the \MSA detector on the prong and the
electrical connections.  The power dissipated in the tuning fork
sensor comes from two sources: the excitation and the vibration of the
tuning fork.  The tuning fork is excited mechanically by a thickness
mode piezocrystal of 1 nF capacitance excited by a voltage of 10
$mV_{rms}$ amplitude, resulting in 2 nW of dissipated power. 
Dissipation due to the vibration of the tuning fork can be estimated
from the current generated by the tuning fork: taking the impedance at
resonance, $R_{res}$=25 K$\Omega$ for our tuning fork, a generated
current of 1 nA gives a dissipated power of 0.025 pW. These values are
compatible with the use of the tuning fork \MSA sensor in a very low
temperature scanning microscope.

\section{REGULATION SCHEME}

The magnetic field due to the vortices spreads rapidly with height. 
Therefore a large scan range is necessary in order to observe a
sufficient number of vortices. As a consequence the microscope needs 
to be much more
flexible than standard AFM or STM microscopes imaging smaller
surfaces. The scanner used \cite {Field} has a scan range of 60 $\mu
m$ at low temperatures for 200 V applied voltage. The eigenfrequency
of the scanner is of the order of 200 Hz, limiting the bandwidth of
regulation. We have adopted a regulation scheme using a phase locked
loop constituted of discrete elements \cite {Rychen_1999}.

\begin{figure}%[h!tb] %h=here,!=au plus prs,t=top,b=bottom
	\includegraphics*[width=7cm]{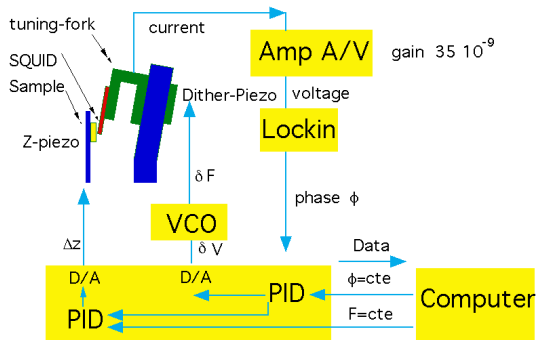}
	\caption{Diagram of the frequency
	and distance regulation used for our SSFM.}
	\label{Regulation}
\end{figure}

The increase of the resonance frequency for 
increasing surface forces is used
for the approach control. A working frequency is chosen 1 Hz higher
than the resonance frequency far away from the sample surface. The 
phase of the tuning fork current is offset to be zero at resonance.  The
tuning fork is excited by a piezoelectric thickness mode plate via a
voltage controlled oscillator (VCO), (Yokogawa FE 200), the excitation
frequency is adjusted continuously using the condition that the phase
must be kept at a fixed value (phase value at resonance).  The
tip-sample distance is fixed by the condition that the excitation
frequency must be equal to the working frequency.  These two
constraints: maintaining the phase locked, and stabilizing the
resonance frequency value, can be satisfied by cascading two
Proportional-Integral (PI) stages, Fig.  \ref{Regulation}.  The first,
maintains the oscillator at resonance in locking the phase, the second
stage uses the deviation of the resonance frequency from the working
frequency to act through a D/A converter on the piezoelectric
z-bender, moving the sample in or out.  The working frequency is set
via a digital input, the analog input takes the phase output of the
digital lockin-amplifier (Perkin Elmer 7280).
%The first PI stage calculates
%the correction necessary to maintain the frequency.  The correction is
%transferred via a D/A converter to the VCO. and the correction serves
%as input for the second PI stage.  This stage calculates the
%correction for the working frequency and the value is transferred via
%a second D/A converter to the z bender.  
The PI stages are programmed
in a PLD device \cite{ALTERA}, driven by a 40 MHz clock. The numeric
PI stages have a frequency bandwidth of about 10 kHz.

\section{PERFORMANCE AND RESULTS}
The \MSA itself is
scanned in close proximity to the sample's surface. 
\MSAs are of such a small size that it is impractical to couple flux to the
\MSA with a flux transformer, a flux locked loop technique can not be
used.  In presence of strong signals the \MSB's critical current varies
over several arches of period $\Phi_{0}$, and the total flux variation
may be obtained by integration of this variation.
% c'est qoui the champ au dessus du fil?

\begin{figure}
 %[h!tb]
 %\centering
 \begin{minipage}{ 4cm}
  \centering 
 \includegraphics[width=4cm]{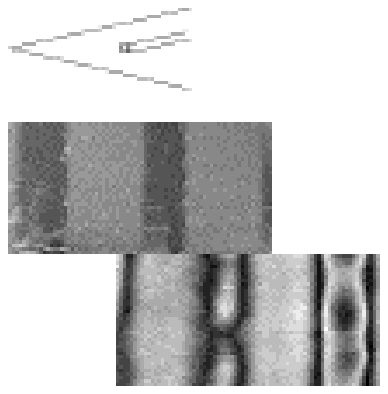}
  %[h!tb] %h=here,!=au plus prs,t=top,b=bottom
 \end {minipage}%
\begin{minipage} {4 cm}
  \centering
  \includegraphics[width=4cm]{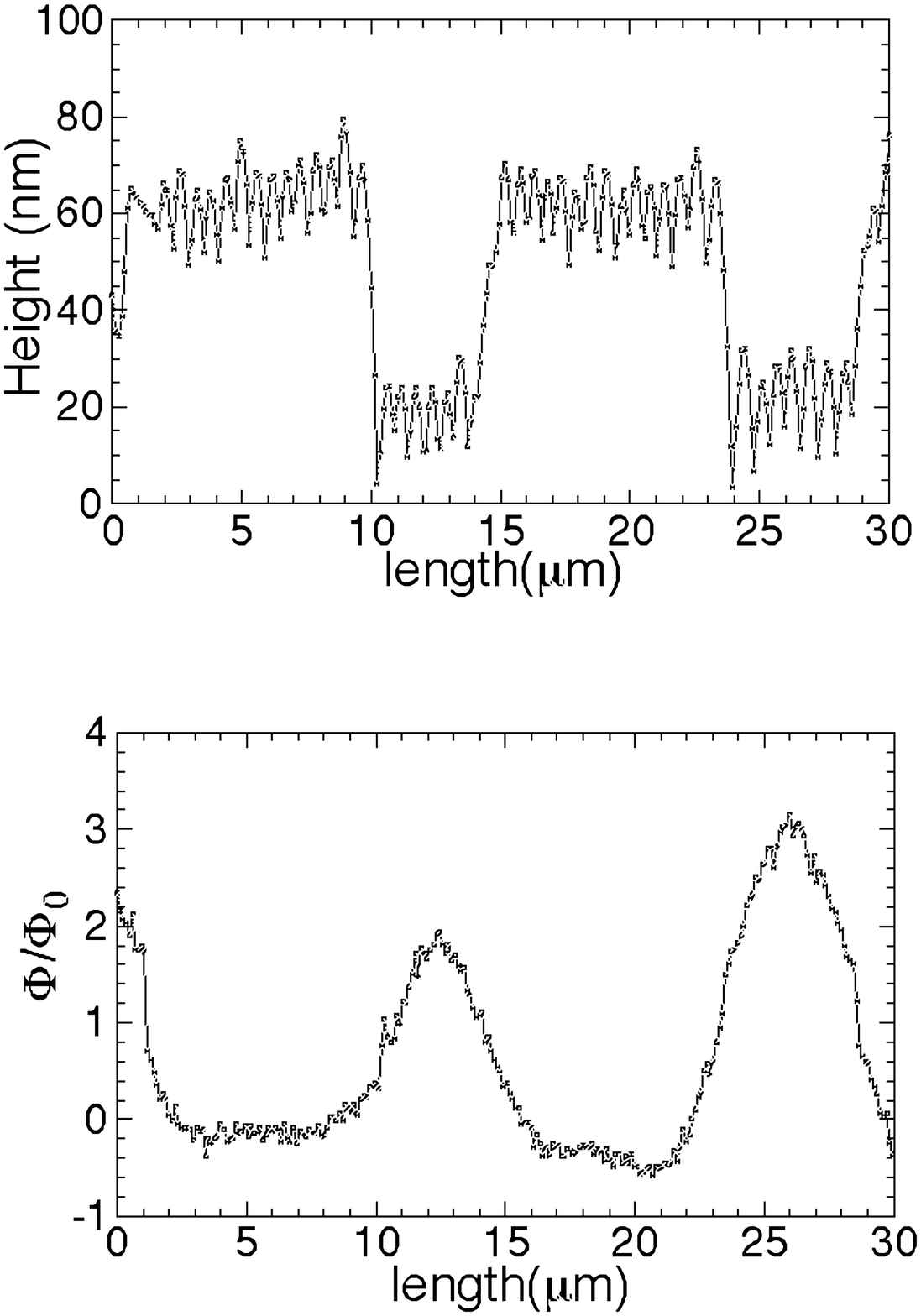}
  %[h!tb] %h=here,!=au plus prs,t=top,b=bottom
 
 \end {minipage}%

% \begin{figure}
%    \vspace{1 cm}
 %\includegraphics*[width=4.cm]{TOPO_MAG_SQUID.eps}
%\includegraphics*[width=4.5cm]{ligne_topo_mag.eps}

  	\caption{left top: Schematic drawing of the \MSA-chip used,
	the magnetic sensor was 15 $\mu m$ away from the tip.  left
  	center: Topographic image of a periodic pattern of 6$\mu m$ wide Nb lines
  	(clear) 8$\mu$m apart, the image is 28 $\mu$m wide and 14 $\mu m$
  	high.  left bottom: Magnetic image taken simultaneously
  	($\mu_{0}H_{ext}$=1 mT, field cooled).  The pictures are laterally
  	offset to reflect the \MSA-tip distance, right top: Line scan through the
  	topographic image, right bottom: Line scan through the magnetic image. 
  	}
	\label{Lignes}
\end{figure}

The geometry of the refrigerator allows thermalization of the
microscope on any stage.  In the first example we used a Nb \MSA patterned
from 30 nm thick film having a
superconducting transition temperature of 7.8 K.
The microscope plate was mounted on the 4.2 K stage.  The sample consisted
of 6 $\mu m$ wide Nb lines 8 $\mu m$ apart.  The lines were obtained
by an RIE process from a UHV e-beam deposited Nb film of 50 nm
thickness.  The sample was mounted on the scanner.  We present
topographic and magnetic images and line scans, Fig. \ref{Lignes}.
The probe was cut with the dicing machine.  The \MSA was
about 15 $\mu $m away from the tip, the angle of approach was
 $\approx$ 10 degrees, resulting in a \MSA-sample distance of 2.6 $\mu
 m$.  For imaging the sample was cooled in a perpendicularly applied
 field of 1 mT and the sample temperature was 5 K. The \MSA temperature
 was regulated at 6.7 K. At this temperature the Nb-\MSAs critical
 current was 400 $\mu A$ with 40 $\mu A$ modulation ($I_{c}$
 ($\Phi_{0}$) - $I_{c}$ ($\Phi_{0}$/2)).

The magnetic image obtained is due to screening currents in the
superconducting lines. The magnetic spatial resolution is limited by the
width of the Nb lines and the \MSA-sample distance. The topographic image
is displaced compared to the magnetic image due to the \MSA-tip
distance Fig.  \ref{Lignes} left top.  On principle the lateral
resolution of the microscope is limited by the shape of the tip: The
sensor cannot probe steep profiles without touching elsewhere, as the
tip is just the apex of the chip and the approach angle is small.  For
example, with an approach angle of 10 degrees and a step height of 50
nm the chip will start to sense the step 300 nm before the apex
reaches the step.

For high resolution measurements we used Al-\MSAs ($T_{c}$=1.2 K) and
placed the microscope plate on the still screen.  A critical current
modulation between 120 $\mu A$ at $\Phi_{ext}$=$\Phi_{0}/2$ and 170
$\mu A$ at $\Phi_{ext}$=$\Phi_{0}$ was obtained at 0.45 K. The \MSA
operating temperature of 0.45 K was reached using silver foil as
thermal link between mixing chamber and the thermally isolated \MSA
head.  The \MSA chip was precision-cut at about 10 $\mu m$ from the
tip, no mesa etch was applied.  The angle of approach was 5 degrees. The
\MSA was used to study a continuous Nb thin film of 200 nm thickness,
evaporated in UHV with an electron beam evaporator and mounted on the
scanner, thermalized at 0.9 K.

\begin{figure}%[h!tb] %h=here,!=au plus prs,t=top,b=bottom
	\includegraphics*[width=8cm]{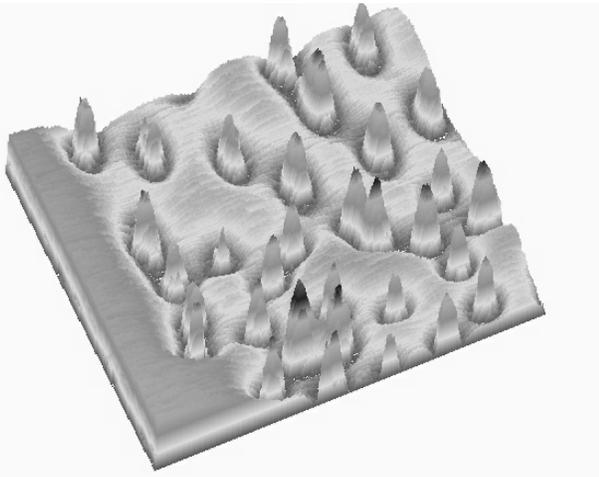}
	\caption{Magnetic flux image of a 200 nm thin Nb film at 0.9 K as measured with
	the Scanning \MSA Force Microscope (SSFM).  Each peak is due to a flux
	quantum produced by a single vortex.  The image size is 28 x 28 $\mu
	m^2$, each pixel has been recorded after averaging 20 \MSA readings.}
	\label{Lattice}
\end{figure}

The \MSA magnetic image, Fig. \ref{Lattice},
shows the spatial dependence of the \MSB's
critical current after field cooling the film in 0.1 mT. The expected
average vortex density for such a field is one vortex in 20.7 $\mu
m^{2}$.  The image size is 28 $\mu m$ a side.  We measure the expected
vortex density.  Each imaged vortex is circled by a line of minimum
critical current.  This is due to the fact that the flux variations
measured causes the \MSB's critical current to pass through a minimum
of the $I_{c}(\Phi)$ characteristic.  The measured $I_{c}(\Phi)$
characteristic can be used to translate the critical current- into the
flux variation.  The Nb film showed strong pinning as the vortex
pattern did not change upon an applied field change of 0.1 mT at low
temperatures in the superconducting state.

\begin{figure}%[h!tb] %h=here,!=au plus prs,t=top,b=bottom
	\includegraphics*[width=7cm]{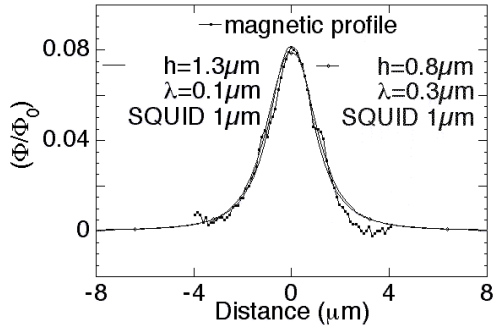}
	\caption{Flux profile of a vortex in the Nb film normalized by the
	quantum of flux of the \MSA. The two different fits represent the
	uncertainty in height due to the uncertainty of \MSA-tip distance.
	The value for the penetration depth, $\lambda$, obtained from
	equation~$\ref{two}$, is strongly dependent on the scan height.  }
	\label{Vortex_fit}
\end{figure}

Fig. \ref{Vortex_fit} shows a line scan through a vortex of Fig. 
\ref{Lattice}. Here we corrected for the periodic $I_{c}(\Phi)$
characteristic. We obtain a well defined bell-shaped curve. The noise
level is about $5 \cdot {10^{-3}} \Phi_{0}$ corresponding to ${10^{-3}}
\Phi_{0}/\sqrt{Hz}$. The width at half height is 2 $\mu m$. This
width is due to the instrumental resolution determined by the distance
between the \MSA and the sample's surface.

A more quantitative mean to analyze one single vortex is to fit the measured
flux profile to the expected flux profile of a single vortex, for a given \MSA
size and \MSA-sample distance.  Nb as a thin film is a type II
superconductor with a normal vortex core diameter of about 20 nm
(coherence length, $\xi$).  In the limit of a small vortex core
diameter, compared to the penetration length and the \MSA-sample
distance, it is possible to use an analytical expression
\cite{John_APL} for the z component of the magnetic field produced by
a vortex.  The field is calculated at a height z above the center of
the film of arbitrary film thickness d, assuming that the core
diameter $\rightarrow 0$.  This approximation is valid as the \MSA
sample distance is of the order of 1$\mu m$, making it impossible for
us to resolve the core of a vortex in Nb.  The expression for the
field is:
\begin{equation}
h_{z}(\mathbf{r},z)= \frac{\Phi_{0}}{(2 \pi \lambda_{ab})^{2} } \int
d^{2}\mathbf{k}\exp^{i\mathbf{kr} } \frac{\exp^{k(d/2-z)}
}{\alpha\lbrack \alpha + k \coth(\alpha d/2) \rbrack}
\label{two}
\end{equation}

where $\mathbf{r}=\lbrace x,y \rbrace$ ,
$\mathbf{k}=\lbrace k_{x},k_{y} \rbrace$,
k=$\sqrt{k_{x}^{2}+k_{y}^{2}}$ and $\alpha$=$\sqrt{k^{2}+\lambda^{-2}
}$.
%\frac{\Phi_{0}{2\pi \lambda_{ab}}$$
 
After integrating the spatial dependence of the predicted magnetic
field over the \MSA's surface it is possible to compare the measured
flux profile with the calculated one.  The adjustable parameters are
$\lambda$ and to a certain degree the \MSA-sample distance.

A height between of 0.8 and 1.3 $\mu m$ corresponding to 9 and 15 $\mu m$ 
\MSA-tip, is consistent with our observations for this particular sensor.  A
penetration depth of 100 to 200 nm is expected for Nb thin films of
good quality of comparable thickness.  The two different fits
presented in Fig.  \ref{Vortex_fit} are limiting cases considering the
uncertainty of the \MSA-sample distance.  They show also how the
\MSA-sample distance intervenes in the interpretation of the
absolute value of the physical quantity reflecting the magnetic field
source (penetration depth).

 \section{DISCUSSION}
 We have built a scanning \MSA force microscope capable of imaging
 individual vortices with high magnetic and spatial resolution.  The
 spatial resolution obtained is the best reported for SQUID
 microscopes. The microscope is based on a combination of the
 lithographically defined \MSA-detector with the force sensor.  The
 force sensor is a quartz tuning fork with a power dissipation in the
 nW range, well adapted for use at low temperatures.  We have
 demonstrated that this microscope has state of the art magnetic
 imaging capabilities combined with topographic imaging at temperatures
 as low as 0.45 K. Improvements of the thermal isolation will allow us to
 image even at lower temperatures.  The spatial and magnetic resolution
 of the \MSA sensor can still be improved.  The mesa etch technique
 demonstrated in Fig.  \ref{pointe} reduces the \MSA sample distance to
 less than 0.5 $\mu m$.  When this short distance is attained the
 spatial resolution \cite{Physica_C} is given by the 200 nm width of
 the \MSA branch.  The sensitivity of the \MSA is limited by the
 measuring scheme, in shunting the \MSA it is possible to increase the
 measuring frequency and thus to increase sensitivity.  
 The microscope
 is working in a dilution refrigerator at 0.45 K, opening new
 possibilities for the direct observation of vortices in unconventional
 $LT_{c}$ superconductors and to answer questions like: How does the
 vortex lattice of $UPt_{3}$ respond upon the crossing of the phase
 boundaries inside the superconducting phase diagram?  Vortex phase
 transitions in microstructured lattices can be studied, leading to the
 understanding of the underlying coherence and frustration effects.

 One key aspect of the microscope is the combination of tuning fork and
 the microfabricated \MSA sensor. This modular approach can be easily
 transferred to other types of microfabricated devices as Hall probes or single
 electron transistor sensors for example.

 \section{ACKNOWLEDGEMENTS}
 The authors thank J.L. Bret and M. Grollier for the building of the
 scanning \MSA electronics, H. Courtois, N. Moussy and M. Bravin for
 sharing their experience in scanning probe microscopy and A. Benoit
 and B. Pannetier for the interest in the ongoing work.  We thank J.
 Kirtley for his help and the fruitful discussions.  This work was
 supported by CNRS ULTIMATECH and DGA contract n$^{\circ}$
 96 136.

%\end{references} 

\end{document}